\documentclass[aps,prl,showpacs,twocolumn,floatfix]{revtex4-1}
\usepackage{graphicx}
\usepackage{amssymb}
\usepackage{subfigure}

\begin{document}

\title{Photoassociation of long-range $nD$ Rydberg molecules}
\author{D.~A.~Anderson*}
\author{S.~A.~Miller}
\author{G.~Raithel}
\affiliation{Department of Physics, University of Michigan, Ann Arbor, MI 48109}

\date{\today }
\begin{abstract}
We observe long-range homonuclear diatomic $nD$ Rydberg molecules photoassociated out of an ultracold gas of $^{87}$Rb atoms for 34$\le n \le$40.  The measured ground-state binding energies of $^{87}$Rb$(nD-5S_{1/2})$ molecular states are larger than those of their $^{87}$Rb$(nS-5S_{1/2})$ counterparts, showing the dependence of the molecular bond on the angular momentum of the Rydberg atom. We exhibit the transition of $^{87}$Rb$(nD-5S_{1/2})$ molecules from a molecular-binding-dominant regime at low $n$ to a fine-structure-dominant regime at high $n$ [akin to Hund's cases (a) and (c), respectively]. In the analysis the fine structure of the $nD$ Rydberg atom and the hyperfine structure of the $5S_{1/2}$ atom are included.
\end{abstract}


\maketitle
Cold atomic systems have opened new frontiers at the interface of atomic and molecular physics.  Of particular interest are a recently discovered class of long-range, homonuclear Rydberg molecules~\cite{Greene.2000,Bendkowsky.2009}.  Formed via an attractive interaction between a Rydberg electron and a ground-state atom~\cite{Greene.2000}, these molecules are among the largest ever observed with internuclear separations of several thousand Bohr radii.  Their distinctive binding mechanism, which is unlike conventional covalent, ionic, and van der Waals bonds between ground-state atoms, results in loosely bound molecules whose properties closely mimic those of their constituent Rydberg atoms.  The discovery of these molecular bonds has been likened to a new ultracold chemistry~\cite{Hogan.2009}, and has spurred a significant amount of theoretical~\cite{Khuskivadze.2002,Sadeghpour.2013} and experimental interest~\cite{Bendkowsky.2010,Li.2011,Tallant.2012,Bellos.2013}.  Non-degenerate, low angular momentum Rydberg states (orbital angular momentum $\ell \le 2$ in rubidium) produce molecules with a few tens of MHz binding energies and permanent electric dipole moments of a few Debye.  The $\ell=0$ molecules were first observed by photoassociation~\cite{Jones.2006} of cold $^{87}$Rb atoms~\cite{Bendkowsky.2009}.  Recently, molecular states with $\ell=1$ and high electron energies have also been excited via bound-bound transitions in $^{85}$Rb$_2$ \cite{Bellos.2013}.  For high-$\ell$ Rydberg states, so-called trilobite molecules with giant permanent electric dipole moments of several kilo-Debye and several GHz binding energies are predicted to exist~\cite{Greene.2000}.  Dipolar Rydberg molecules have previously been prepared in both Rb~\cite{Li.2011} and Cs~\cite{Tallant.2012}.

The relevant interaction was first described by Fermi \cite{Fermi.1934} to help explain pressure-induced energy shifts of Rydberg absorption lines in a gas~\cite{Amaldi.1934}.  The deBroglie wavelength of the Rydberg electron (position ${\bf{r}}$) is much larger than that of a heavy ground-state atom (position ${\bf{R}}$) that lies within the Rydberg atom's volume, and their interaction can be approximated as a low-energy s-wave scattering process (scattering length $a_s$). The interaction is described with a Fermi-type pseudopotential~\cite{Omont.1977,Greene.2000},
$V_{\rm{pseudo}}({\bf{r}})=2\pi a_s \, \delta^3({\bf{r}}-{\bf{R}} )$,
where p-wave and higher-order scattering~\cite{Khuskivadze.2002} are neglected. For negative $a_{s}$ the interaction can lead to bound molecular states \cite{Omont.1977,Greene.2000}.

In the present work we focus on long-range $^{87}$Rb$_2$ molecules formed by an $nD$ Rydberg and a $5S_{1/2}$ ground state atom. The binding energies generally increase with $\ell$, due to the $\sqrt{2 \ell +1}$-scaling of the $Y_l^{m=0}(\theta=0)$. Among the low-$\ell$ variety of these molecules the $nD$ ones have the highest binding energies. The angular-momentum coupling spans three Hund's cases when varying $\ell$ from 0 to 2. The $nS_{1/2}-5S_{1/2}$ molecules are akin to Hund's case (b), because they have $L=0$ and total electron spin $S=1$. The $nP_j-5S_{1/2}$ molecules are akin to Hund's case (c), because the fine structure strongly dominates the molecular binding. The $nD-5S_{1/2}$ molecules range from Hund's case (c) for large $n$ to Hund's case (a) for $n \lesssim 30$, where the molecular binding increasingly dominates the fine structure. The $F$- and higher-$\ell$ states of Rb intersect with a sole, deeply bound potential (the trilobite potential) which accumulates most of the level shifts~\cite{Greene.2000}. This negates the existence of molecular potentials wells for $F$- and higher-$\ell$ states similar to those below the $S$, $P$ and $D$ asymptotes.

To excite $nD$ Rydberg molecules we first prepare a sample of $\sim$10$^{5}$ magnetically trapped $^{87}$Rb atoms in their $|F$=2, $m$$_{F}$=2$\rangle$ ground state at a temperature $\le$17$~\mu$K and peak density $\gtrsim5\times10^{11}$~cm$^{-3}$.  Optical excitation to $nD$ Rydberg states is accomplished via a two-photon transition from the 5$S_{1/2}$ ground state using 780~nm and 480~nm laser beams. The 780~nm laser frequency is fixed $\sim$1~GHz off-resonance from the 5$S_{1/2}$ to 5$P_{3/2}$ transition to avoid atom heating, and the 480~nm laser frequency is scanned to excite either Rydberg atoms or Rydberg molecules.  The combined excitation bandwidth is $\approx$2~MHz.  The 780~nm laser has a power of $\sim$500~$\mu$W and is collimated to a full-width half-maximum (FWHM) of 3.5~mm. The 480~nm beam has a power of $\sim$35~mW and is focused to a FWHM of $89\pm 5~\mu$m into the cigar-shaped atom sample, which  has a FWHM diameter of $28~\mu$m and an aspect ratio of $\approx 1:3$.

The atom sample is enclosed by six individually-addressable electrodes that are used to control the electric field at the excitation location, described in previous work~\cite{Anderson.2013}.  We zero the electric field in this region to within $\lesssim$200~mV/cm in all coordinates by Stark spectroscopy on $59D$ Rydberg states~\cite{Neukammer.1987}.  This ensures that residual quadratic Stark shifts of the $nD$ Rydberg levels in the $n$ range of interest are $\lesssim$2~MHz.  In a single experiment, the ground-state atom sample is illuminated by 2-3~$\mu$s long laser pulses followed by electric-field ionization of Rydberg atoms and molecules~\cite{Gallagher}. The signal ions are extracted and detected by a micro-channel plate located 10~cm from the excitation region.  We use one ground-state atom sample for a series of 55 individual experiments at a single 480~nm frequency step.  The loss of phase-space density of the ground-state atom sample over the course of an excitation series is insignificant.
\begin{figure}[h]
\includegraphics[width=8.7cm]{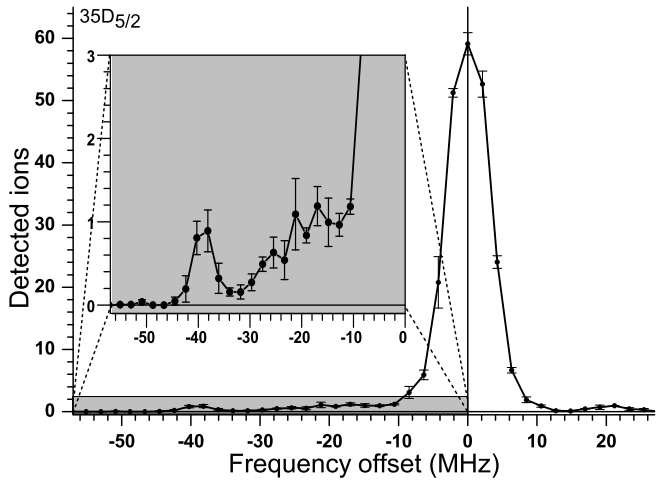}
\caption{Spectrum centered on the $35D_{5/2}$ atomic Rydberg line showing the $^{87}$Rb$(35D_{5/2}-5S_{1/2})(\nu=0)$ molecular line at $-38 \pm 3$~MHz.  The vertical error bars are the standard error of 3 sets of 55 individual experiments at each frequency step.  The error in the binding energy is estimated to be $\pm 3$~MHz, limited by the average long-term frequency drift observed over one full scan.}
\label{fig:35d2pt5}
\end{figure}

The photoassociation of a Rydberg atom and ground-state atom pair into a bound molecular state occurs when the excitation laser is detuned from the atomic Rydberg line by an amount equal to the molecular binding energy.  In Fig.~\ref{fig:35d2pt5} we show an experimental spectrum in the vicinity of the atomic 35$D_{5/2}$ Rydberg line.  A prominent satellite line emerges at $-38\pm3$~MHz, which is assigned to the $(35D_{5/2}-5S_{1/2})(\nu=0)$ molecule, where $\nu=0$ denotes the vibrational ground state.  This binding energy is $\approx 1.6$ times larger than that of the $(35S_{1/2}-5S_{1/2})(\nu=0)$ molecular ground state measured in previous experiments~\cite{Bendkowsky.2009,Bendkowsky.2010}.  The larger binding energy of the $nD$ molecule reflects the expected deepening of the molecular potential with $\ell$.

\begin{figure}[h]
\includegraphics[width=8.7cm]{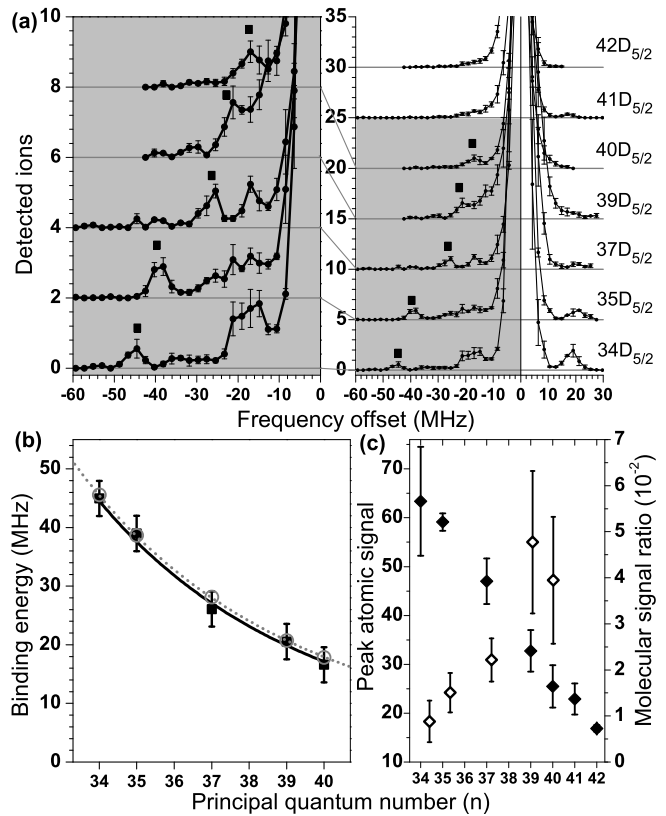}
\caption{(a) Right: Spectra centered on $nD_{5/2}$ atomic Rydberg lines for the indicated values of $n$ and identified molecular lines (squares). Left: Selected spectra from plot on right for states with identified molecular lines.  Error bars are obtained as in Fig.~\ref{fig:35d2pt5}.  (b) Binding energies obtained from Gaussian fits to the molecular lines identified in (a) vs $n$.  An allometric fit (solid curve) to the experimental binding energies yields a $n^{-5.9\pm0.4}$ scaling.  Also shown are theoretical binding energies for the $^{87}$Rb$(nD_{5/2}-5S_{1/2})(\nu=0)$ molecular states with $a_{s0}$=$-14~a_0$ (hollow circles and dotted curve).  (c) Peak number of detected ions on the $nD_{5/2}$ atomic Rydberg line (solid diamonds, left axis) and ratio of molecular and atomic line strengths (hollow diamonds, right axis) vs $n$.}
\label{fig:2pt5_waterfall}
\end{figure}

A series of $nD_{5/2}$ Rydberg spectra in the range $34~\le~n~\le~42$ is shown in Fig.~\ref{fig:2pt5_waterfall}a (right).  The lowermost, red-shifted lines for $n$=40 and below are assigned to the molecular states.  Molecular lines are not discernable in the $n$=42 and 41 spectra because the line broadening due to residual fields and the laser linewidth exceeds the molecular binding energies for these states. Additional satellite lines with smaller binding energies are not unexpected, but are likely obscured in Figs.~\ref{fig:35d2pt5} and~\ref{fig:2pt5_waterfall}a by the broadening of the atomic lines as well as artificial signals at $\pm 20~$MHz due to weak, symmetric side peaks in the 480nm laser spectrum (caused by a Pound-Drever-Hall stabilization loop). Features near $-20~$MHz can only be assigned to molecular lines if they are significantly stronger than the artificial signal seen at $+20~$MHz.

Fig.~\ref{fig:2pt5_waterfall}b shows the molecular binding energies measured in Fig.~\ref{fig:2pt5_waterfall}a vs $n$.  One may expect the binding energies to be proportional to the probability density of the Rydberg electron wavefunction, which scales as $\sim n^{-6}$.  An allometric fit to the data in Fig.~\ref{fig:2pt5_waterfall}b qualitatively supports this expectation over the displayed range of $n$.

As shown in Fig.~\ref{fig:2pt5_waterfall}c, the ratio of molecular and atomic line strengths ranges from $\approx 1-5\%$. Taking our peak atomic density into consideration, this agrees quite well with the relative signal strengths found in~\cite{Bendkowsky.2009,Bendkowsky.2010}.  One may expect the molecular-signal ratio to scale with the probability of finding a $5S_{1/2}$ atom within the Rydberg-atom volume (which scales as $n^6$), corresponding to an increase by about a factor of 2.5 over the $n$-range in Fig.~\ref{fig:2pt5_waterfall}c. The actually observed increase is about a factor of 5. The enhanced increase is likely due to a Rydberg excitation blockade caused by electrostatic Rydberg-atom interactions~\cite{Reinhard.2007, Schwarzkopf.2011}, which suppresses the atomic line~\cite{Tong.2004}. Since the blockade's effectiveness increases with $n$, the molecular-signal ratio scales faster than $n^6$. This interpretation is corroborated by the  atomic-signal strength, which drops by a factor of 4 over the $n$-range in Fig.~\ref{fig:2pt5_waterfall}c. In the absence of an excitation blockade, the atomic signal would drop as $n^{-3}$, {\sl i.e.} by only a factor of 2 in Fig.~\ref{fig:2pt5_waterfall}c.

\begin{figure}[h]
\includegraphics[width=8.5cm]{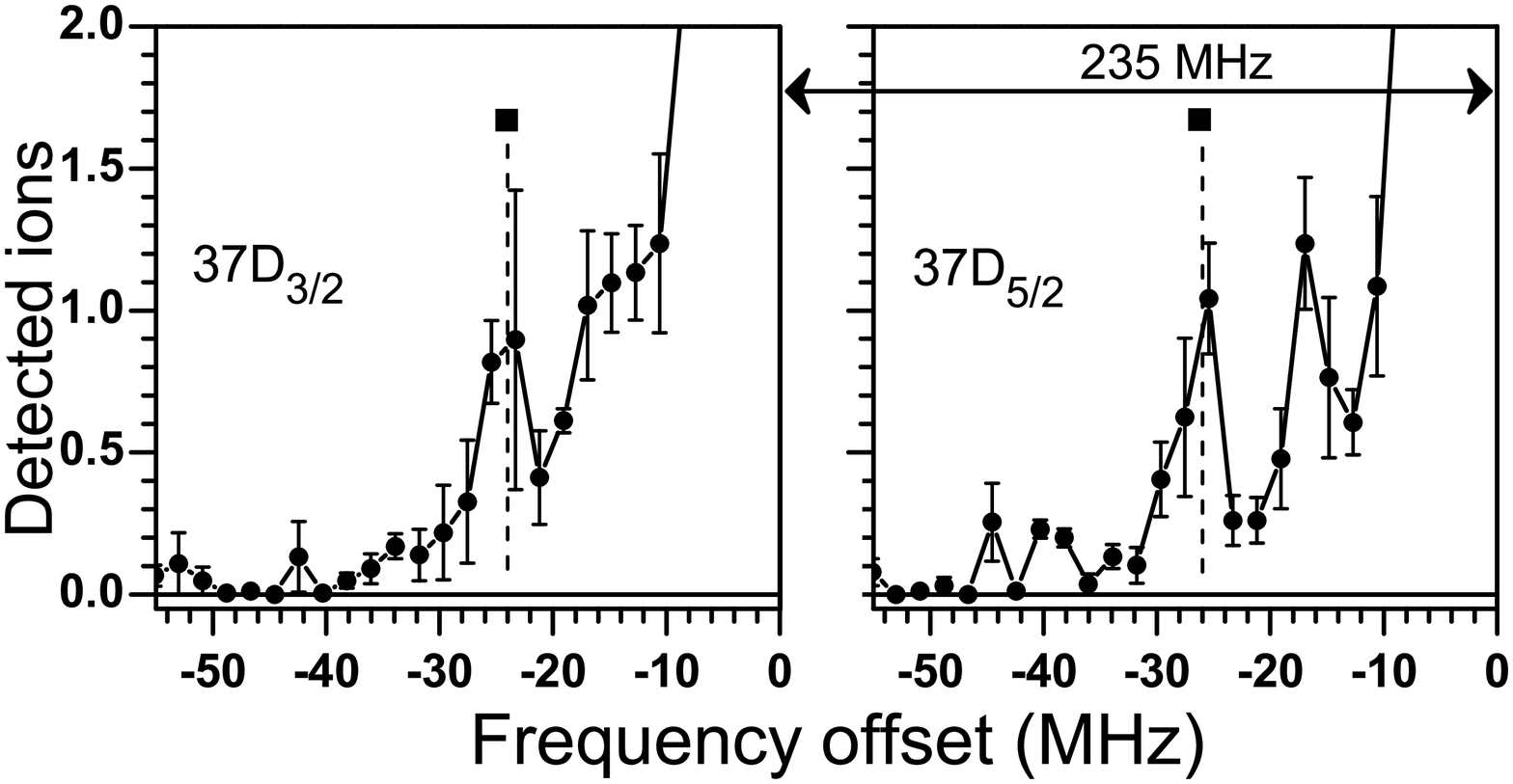}
\caption{Spectra centered on the $37D_j$ atomic Rydberg lines for $j={3/2}$ (left) and $j={5/2}$ (right; same as in Fig.~\ref{fig:2pt5_waterfall}).  Error bars are obtained as in Fig.~\ref{fig:35d2pt5}.  $^{87}$Rb$(37D_{j}-5S_{1/2})(\nu=0)$ molecular signals are indicted by vertical dashed lines and squares.}
\label{fig:37d1pt5}
\end{figure}

Molecular states of the $nD-5S_{1/2}$-type exhibit a transition between Hund's case (a) at low $n$ and Hund's case (c) at high $n$. Most of our data trend towards Hund's case (c), where the dominant molecular potentials carry spin-dependent factors $\vert \langle m_{\ell} = 0, m_{s}= \pm 1/2 \vert j, m_j= \pm 1/2 \rangle \vert^2$ (detailed model below); these factors are $\ell / (2 \ell +1)$ for $j=\ell-1/2$ and $(\ell+1) / (2 \ell +1)$ for $j=\ell+1/2$. Hence, in Hund's case (c) (high $n$) the $\nu=0$ binding-energy ratio for $nD_{3/2}-5S_{1/2}$ and $nD_{5/2}-5S_{1/2}$ molecules should be about 2/3.  For decreasing $n$, the fine structure splitting increases as $n^{-3}$ while the molecular binding energy increases as $n^{-6}$. The system then trends towards Hund's case (a),
in which the $\nu=0$ binding energy ratio changes from 2/3 to $\gg 1$ (see below). Therefore, the $\nu=0$ binding energy ratio is a convenient experimental measure to characterize the system. Figure~\ref{fig:37d1pt5} shows spectra of the two fine structure components of $37D$.  Molecular peaks are present for both $j=3/2$ and $5/2$, with respective binding energies of
$24 \pm 3$~MHz and $26 \pm 3$~MHz, corresponding to a ratio of $0.92 \pm 0.15$. Since this is significantly larger than 2/3, for $n=37$ the system is in the transition regime between Hund's case (c) and (a).

Next we compare experimental data with a calculation.  Triplet scattering leads to molecular binding, while singlet scattering is not relevant because it merely generates weak, repulsive potentials~\cite{Khuskivadze.2002}. For low electron momenta, $k$, the s-wave triplet scattering length $a_s^T(k) \approx a_{s0}+\frac{\pi}{3}\alpha k$, where $a_{s0}=a_{s}^{T}(k=0)$ is the zero-energy scattering length and $\alpha$ the Rb~$5S_{1/2}$ polarizability~\cite{Omont.1977}.  We neglect p-wave scattering and fit $a_{s0}$ to match the experimental data. The Rydberg atom's fine structure and the perturber's hyperfine structure are included because they are on the same order or larger than the molecular binding. The much smaller hyperfine structure of the Rydberg atom is ignored.  For the $5S_{1/2}$ atom located at ${\bf{R}} = Z {\bf{e}}_z$, the Hamiltonian is
\begin{equation}
\hat{H}_0 + 2 \pi a_{s}^T (k(r)) \delta^3 (\hat{{\bf r}} - Z {\bf e}_z) (\hat{\bf S}_1 \cdot \hat{\bf S}_2 + \frac{3}{4}) + A \hat{\bf S}_2 \cdot \hat{\bf I}_2
\label{Had}
\end{equation}
where the unperturbed Hamiltonian $\hat{H}_0$ includes Rydberg quantum defects and fine structure~\cite{Gallagher}. The operators $\hat{\bf S}_1$ and $\hat{\bf S}_2$ are the spins of the Rydberg electron and $5S_{1/2}$ atom, respectively. The $^{87}$Rb~$5S_{1/2}$ atom has $\ell_2=0$, a nuclear spin $\hat{\bf I}_2$ with $I_2=3/2$, and a hyperfine parameter $A = h \times 3.4$~GHz. The projector $\hat{\bf S}_1 \cdot \hat{\bf S}_2 + \frac{3}{4}$ has the eigenvalue one (zero) for the triplet (singlet) states of $\hat{\bf S}_1$ and $\hat{\bf S}_2$; it thus enables only triplet scattering. In the classically allowed range of the Rydberg electron $k=\sqrt{-1/(n_{\rm eff,1} \, n_{\rm eff,2})+2/r}$ (at. un.), and $k=0$ elsewhere. There, $n_{\rm eff,1}$ and $n_{\rm eff,2}$ are the effective quantum numbers of the Rydberg states coupled by the scattering term. Since only states with $m_{\ell 1}=0$ are non-vanishing on the internuclear axis, the relevant Hilbert space is limited to $\{ \vert n, \ell_1, j_1 , m_{j1}=\pm 1/2, m_{s2}= \pm 1/2, m_{i2}=\pm 1/2, \pm3/2 \rangle \}$. Since the Hamiltonian in Eq.~\ref{Had} conserves $m_k := m_{j1} + m_{s2} + m_{i2}$, the space breaks up into separated sub-spaces with $m_k=\pm 5/2, \pm 3/2, \pm 1/2$, within which the scattering term couples states with same $m_{j1}+m_{s2}$, while the hyperfine term couples states with same $m_{s2}+m_{i2}$.

The Hamiltonian in Eq.~\ref{Had} is diagonalized, resulting in adiabatic potential surfaces $V_{ad,i}(Z)$, and electric ($d_{i}(Z)$) and magnetic ($\mu_{i}(Z)$) dipole moments of the adiabatic states ($i$ is an arbitrary label for the $V_{ad}(Z)$). The vibrational states $W_{i,\nu}$ and their wavefunctions $\Psi_{i,\nu}(Z)$ are found by solving the Schr{\"o}dinger equation with potential $V_{ad,i}(Z)$ and reduced mass 87~amu/2, and their electric and magnetic dipole moments are $d_{i,\nu} = \int \vert \Psi_{i,\nu}(Z)\vert^2 d_{i}(Z) dZ$ and $\mu_{i,\nu} = \int \vert \Psi_{i,\nu}(Z)\vert^2 \mu_{i}(Z) dZ$.

\begin{figure}[h]
\includegraphics[width=8.5cm]{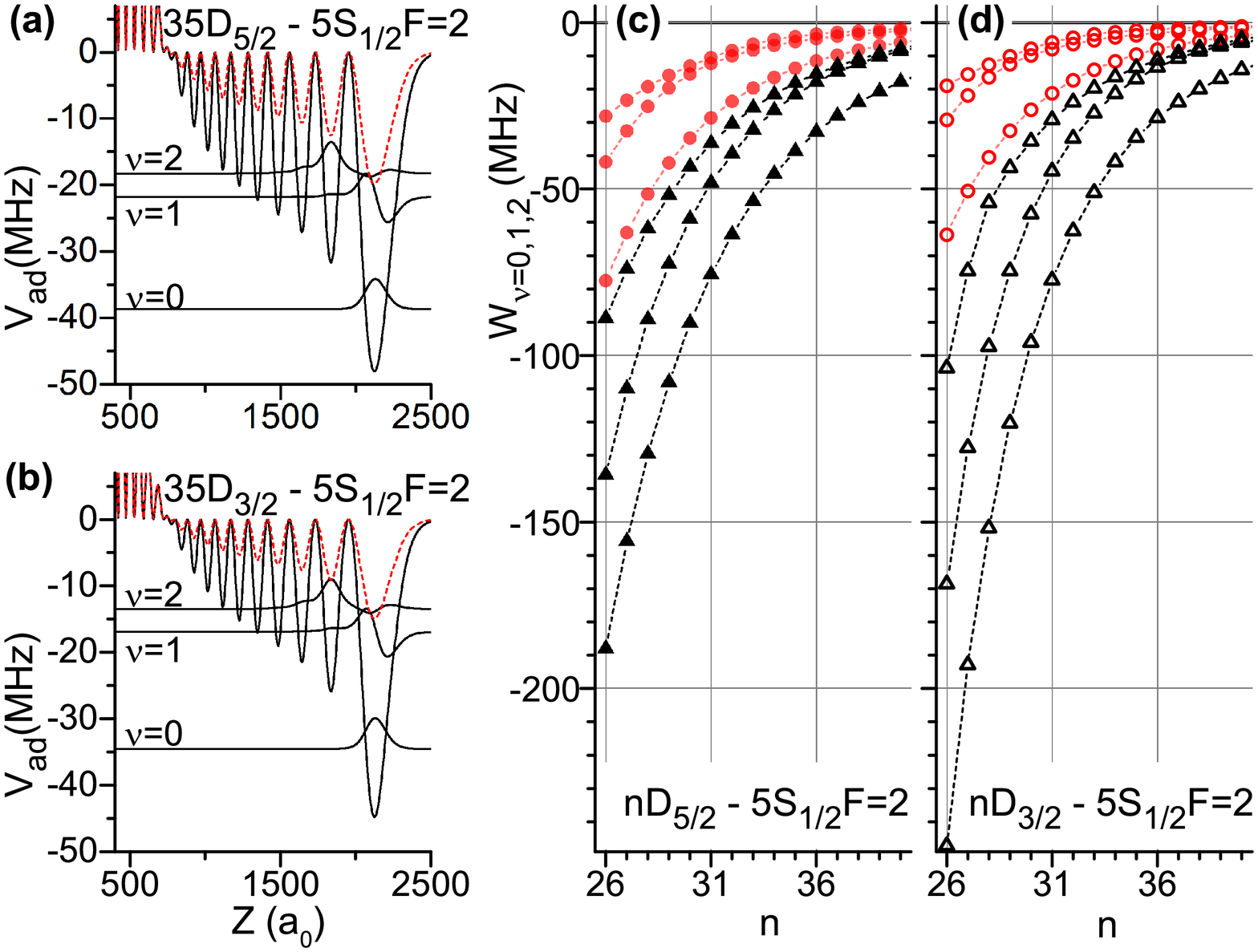}
\caption{(color online).  (a) and (b): Deep (solid) and shallow (dashed) adiabatic potentials for $35D_{3/2}$ and $35D_{5/2}$-type molecules, for $a_{s0}=-14~a_0$, and vibrational wavefunctions for $\nu =0,1,2$ in the deep potentials. (c) and (d): Energy levels $W_{i,\nu}$ for $\nu =0,1,2$ in the deep (triangles) and shallow (circles) potentials vs $n$.}
\label{fig:theory1}
\end{figure}

In Fig.~\ref{fig:theory1}a and b we show all potentials $V_{ad}(Z)$ for $35D_{5/2}$ and $35D_{3/2}$ that connect to the $F=2$ hyperfine level of the $5S_{1/2}$ atom, as well as the vibrational states $\nu =0,1,2$ of the deeper potentials. The deep and shallow potentials and their states have degeneracies of 6 and 4, respectively. In Fig.~\ref{fig:theory1}c and d we show the vibrational energies $W_{i,\nu}$ for $\nu = 0,1,2$ over a range of $n$ for $nD_{5/2}-5S_{1/2}$ and  $nD_{3/2}-5S_{1/2}$ molecules.

In the experiment we excite molecules below the $F=2$ asymptote, for which the states in the deep potential $V_{ad}(Z)$ have the larger degeneracy and are easier to observe due to their larger binding energies. Hence, we adjust $a_{s0}$ so that the $\nu=0$ levels in the deep $V_{ad}(Z)$ match the experimental data in Figs.~1 to~3, and find $a_{s0}=-14~a_0 \pm 0.5~a_0$ (see Fig.~\ref{fig:2pt5_waterfall}b). (For 37$D_{5/2}$ the $\nu = 0$ binding energy increases by about 4~MHz when changing $a_{s0}$ from $-13.5$ to $-14.5~a_0$.) This $a_{s0}$-value lies within the range of published values $-13$ to -19.48~$a_0$~\cite{Bahrim.2000,Bahrim.2001,Fabrikant.1986,Bendkowsky.2009,Bendkowsky.2010}. The experimental signals at about half the $\nu = 0$ binding energies, seen in some of the plots in Fig.~\ref{fig:2pt5_waterfall}a, may correspond to a cluster of three levels at about half the $\nu=0$ binding energies in Fig.~\ref{fig:theory1}c and d.

\begin{figure}[h]
\includegraphics[width=8.5cm]{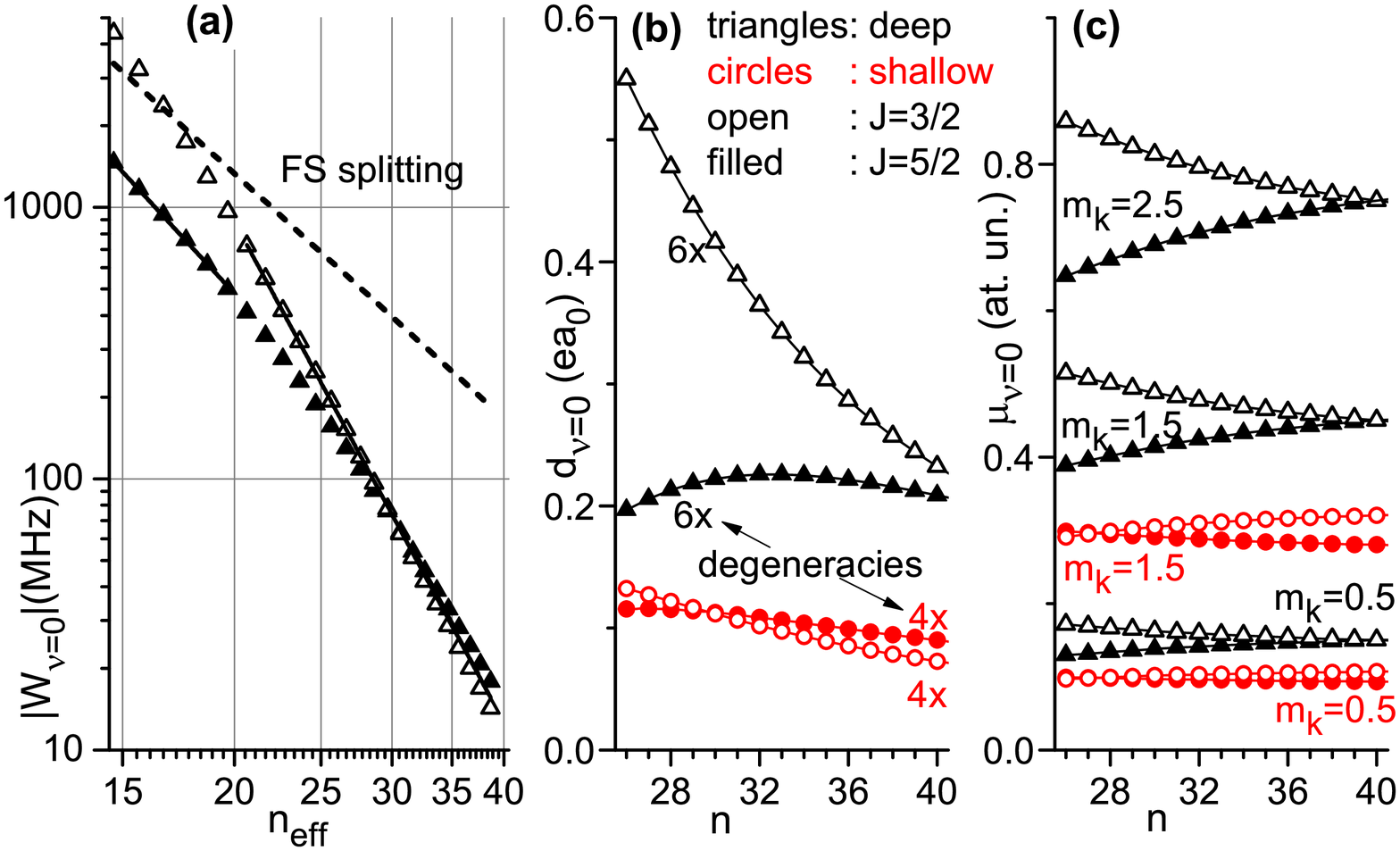}
\caption{(color online).  (a) $\nu=0$ binding energies in the deep molecular potentials for $nD_{5/2}$ (filled triangles), $nD_{3/2}$ (open triangles), and the $D$ fine structure splitting (dashed line), vs effective quantum number. Solid lines are fits. The $D_{3/2}$ energies are fit well by 84.2~GHz/$n_{\rm eff}^{6.13}$. The $D_{5/2}$ energies do not exhibit a global scaling; at low $n$ they tend to scale as 23~MHz/$n_{\rm eff}^{3.6}$. (b) Electric dipole moments for $\nu = 0$ vs $n$ for the deep (triangles) and shallow (circles) $V_{\rm ad}(Z)$ for $j=3/2$ (open) and $j=5/2$ (filled). (c) Magnetic dipole moments for the same states as in (b).}
\label{fig:theory2}
\end{figure}

In the high-$n$ limit in Fig.~\ref{fig:theory2}a the $\nu=0$ binding-energy ratio for $D_{3/2}$ and $D_{5/2}$ approaches $2/3$, as expected for Hund's case (c), and both sets of binding energies approximately scale inversely with the atomic volume (i.e., as $n_{\rm eff}^{-6})$). At low $n$, the system transitions into Hund's case (a). There, the binding-energy ratio changes from $2/3$ to $\gg1$.  Also, the $\nu=0$ binding energies for the lower ($D_{3/2}$) fine structure level exceed the fine structure coupling and keep scaling as $n_{\rm eff}^{-6}$, while those for the $D_{5/2}$ level approach the fine structure splitting and its scaling ($n_{\rm eff}^{-3}$).
The experimental data in Fig.~\ref{fig:37d1pt5} are in the transition regime between the two Hund's cases.

For $D$-type Rydberg molecules the electric dipole moments $d_{\nu=0}$ are somewhat smaller than for $S$-type ones (see Fig.~\ref{fig:theory2}b and Ref.~\cite{Li.2011}). As a result of the transition between two Hund's cases, the $d_{\nu=0}$ of the $nD_{5/2}$-type molecules do not exhibit a clear scaling. Those for the $nD_{3/2}$-type molecules scale close to $n_{\rm eff}^{-2}$, similar to $S$-type ones~\cite{Li.2011}.  Noting that $d_{\nu=0} \lesssim 0.25~ea_0$ and that the stray electric field in our experiment is below 200~mV/cm, the experimental permanent-electric-dipole shift is well below 1~MHz. The magnetic moments (see Fig.~\ref{fig:theory2}c) exhibit a gradual change that reflects the change in angular-momentum coupling behavior between the two Hund's cases. In our $<1$ Gauss magnetic field in the atom trap, the molecular lines could be broadened over a full range of $\lesssim 4$~MHz, while the atomic lines have $\vert m_j \vert \le 5/2$ and could be broadened over a range of $\lesssim 8$~MHz. These estimates correspond quite well with the linewidths seen in the experimental data.

In summary, we have observed polar Rydberg molecules of the type $nD_j - 5S_{1/2}$. The binding energies are larger than those of previously observed  $nS_{1/2} - 5S_{1/2}$ molecules. The molecules undergo a transition between Hund's case (a) at low $n$ to Hund's case (c) at high $n$. With improved spectroscopic resolution one should be able to measure the electric dipole moments, similar to previous experiments with $S$-type molecules~\cite{Li.2011}, and to assign higher-lying vibrational levels in the deep potentials in Fig.~\ref{fig:theory1} as well as levels in the shallow potentials.

This work was supported by the AFOSR (FA9550-10-1-0453).  We thank C. H. Greene, H. R. Sadeghpour and S. T. Rittenhouse for helpful discussions.

*anderda@umich.edu

\end{document}